# Effect of heat treatments and coatings on the outgassing rate of stainless steel chambers

Md Abdullah A. Mamun and Abdelmageed A. Elmustafa

*Dept. of Mechanical and Aerospace Engineering, Old Dominion University, Norfolk, Virginia 23529*

Marcy L. Stutzman[a], Philip A. Adderley, and Matthew Poelker

*Thomas Jefferson National Accelerator Facility, Newport News, Virginia 23606*

[a] Electronic mail: marcy@jlab.org

## ABSTRACT

The outgassing rates of three nominally identical 304L stainless steel vacuum chambers were measured to determine the effect of chamber coatings and heat treatments. One chamber was coated with titanium nitride (TiN) and one with amorphous silicon (a-Si) immediately following fabrication. The last chamber was first tested without any coating, and then coated with a-Si following a series of heat treatments. The outgassing rate of each chamber was measured at room temperatures between 15 and 30 °C following bakes at temperatures between 90 and 400 °C. Measurements for bare steel showed a significant reduction in the outgassing rate by nearly a factor of 20 after a 400 °C heat treatment ($3.5\times10^{-12}$ Torr L $s^{-1}$ $cm^{-2}$ prior to heat treatment, reduced to $1.7\times10^{-13}$ Torr L $s^{-1}$ $cm^{-2}$ following heat treatment). The chambers that were coated with a-Si showed minimal change in outgassing rates with heat treatment, though an outgassing rate reduced by heat treatments prior to a-Si coating was successfully preserved throughout a series of bakes. The TiN coated chamber exhibited remarkably low outgassing rates, up to four orders of magnitude lower than the uncoated stainless steel, but the uncertainty in these rates is large due to the sensitivity limitations of the spinning rotor gauge accumulation measurement and the possibility of a small pump speed due to inhomogeneity in the TiN coating. The outgassing results are discussed in terms of diffusion-limited versus recombination-limited processes.

# I. INTRODUCTION

The successful operation of a GaAs-based spin-polarized electron source[1,2] requires vacuum near $1\times10^{-12}$ Torr [(ultra-high vacuum (UHV) to extreme-high vacuum(XHV)[3]] since residual gasses in the system are ionized by the electron beam and accelerated into the photocathode, causing damage and limiting the operational lifetime of the system. Proposed accelerator electron sources require higher current and corresponding vacuum improvements. The equilibrium pressure in the system, $P$, is defined by the equation, $P = Q/S$ where $Q$ is the gas load from all sources including outgassing and leaks, and $S$ is the effective pump speed. To reduce pressure, the pump speed should be maximized and the outgassing rate minimized[3,4]. For stainless steel systems with no leaks and where no process gas is introduced, the gas load is primarily from hydrogen outgassing from the steel[5]. There are numerous reports of successful reduction of the outgassing rate via moderate[6-12] and high temperature[11-18] heat treatments which serve to reduce hydrogen outgassing. There are also numerous reports of coatings such as titanium nitride (TiN)[19-24], boron nitride[24,25], silicon[26,27], or silicon oxide[28] that may reduce the outgassing rate by acting as diffusion barriers. The experimental results presented here reproduce the outgassing rate reduction for moderate heat treatment temperatures of stainless steel, and represent the first test of the amorphous silicon coating for baked vacuum chambers. Experimental results presented here for TiN coated chambers replicate the extremely low measured outgassing rates in the literature for TiN coated chambers but with large experimental uncertainty in the measurements, and indication of a potential pump speed in the coating which would invalidate the accumulation method of measuring outgassing rate.

These experiments were conducted by measuring outgassing from complete vacuum chambers that included welds and flanges to provide an outgassing rate representative of what can be achieved for production vacuum chambers for electron sources at Jefferson Lab. Outgassing rates were measured following each of a series of system bakes, and at a range of room temperatures. The room temperature dependence of the hydrogen outgassing rate allows measurement of the activation energy for hydrogen diffusion in the bare steel system, and is the basis for the discussion of diffusion versus recombination limitations in the outgassing process.

Scanning electron microscopy (SEM) and surface profilometry were used to determine the topography and mechanical roughness of the coated and uncoated surfaces, and energy-dispersive x-ray spectroscopy (EDS) was used to assess the chemical composition of the coatings.

# II. EXPERIMENT

## *A. Chamber fabrication*

Each vacuum chamber, 20.3 cm in diameter and 38.1 cm long, was manufactured from 304L stainless steel sheet, 0.48 cm thick with a 0.8 μm RMS surface finish that was rolled into a cylinder and welded. A DN200 Conflat™ flange was welded to one end, and a flat plate of the same thickness and material as the walls was welded to the bottom. The chamber was assembled (see Fig.1) with either a zero-length DN200 to DN35 reducer flange, or a custom five-port DN200 adapter flange, thinned to 4.78 mm, with five DN35 ports to support a subsequent experiment. DN35 Conflat™ tees were used to attach a spinning rotor gauge (SRG[29]) and an all-metal right-angle isolation valve (400 °C bakable) to each chamber. Each tee was heat treated and coated identically to the corresponding test chamber, and the dimensions of the tee were included in calculations of the geometric surface area and volume. The additional surface area due to microscopic roughness was not included in calculations of outgassing rate, and the chambers were not electropolished. Chamber volumes and areas were nominally 11.3 L and 3200 $cm^2$, respectively, with exact dimensions used for each system and uncertainties estimated to be below 2%.

After manufacture, all chambers were cleaned with a detergent[30] and solvents, then evacuated and leak checked using a residual gas analyzer (RGA) with a detectable leak threshold below $5\times10^{-11}$ Torr L $s^{-1}$.

Two chambers were coated immediately following fabrication with either titanium nitride or amorphous silicon, and are referred to as TiN and a-Si respectively. The stainless steel chamber was evaluated initially without any coating, and is referred to as SS1. Following a series of heat treatments, SS1 was sent to be coated with the a-Si coating, and is referred to as SS1:a-Si following the coating process. Additional outgassing data from previously published outgassing measurements of a stainless steel chamber[2] is included as SS2 in Fig. 7.

Prior to measuring outgassing rates, all chambers were evacuated and baked in a hot air oven. Following each chamber bake, outgassing rates were evaluated at room temperatures ranging from 15 to 30 °C in a clean room with temperatures stable to within 1 C°. For 400 °C heat treatments, the ion pump and RGA were located outside the oven, and they were baked to at least 150 °C using resistive heat tapes for the duration of the heat treatment, while during bakes at 250 °C and below, the ion pump and RGA were within the hot air oven.

## B. Coating details

The commercial coating SilcoGuard®-1000 from the vendor SilcoTek Corporation[31] is an inert multi-layered barrier of amorphous silicon inter-diffused with the substrate resulting in a 400-800 nm coating. The silicon coating is applied by chemical vapor deposition at approximately 400 °C. The present study was designed to evaluate its potential use in baked systems for UHV and XHV applications.

The TiN coating was provided by Ionbond LLC[32] using cathodic arc physical vapor deposition. The coating was between 9 and 10 µm thick, which is considerably thicker than other coatings in literature for UHV applications (for example, those in refs. 19, and 20, which are between 200 nm and 1 µm).

## C. Outgassing measurement

The rate of rise, or accumulation, technique[33] with the SRG was used to measure the outgassing rate of each test chamber. This technique relies on the measurement of the pressure inside the isolated test chamber as a function of time, $dP/dt$, with the outgassing rate, $q$, given by the expression:

$$q = \frac{dP}{dt} \times \frac{V}{A} \left( \frac{\mathrm{Torr} \cdot \mathrm{L}}{\mathrm{s} \cdot \mathrm{cm}^2} \right) \tag{1}$$

where $V$ is the chamber volume, and $A$ is the chamber surface area. The SRG is a direct reading gauge, and was used because it disturbs the vacuum minimally compared to ionization gauges, which can generate a gas load and/or provide a small amount of pumping depending on system pressure. The SRG lower measurement limit is approximately $1\times10^{-7}$ Torr.

Outgassing rates were calculated using pressure measurements obtained over a period of at least 24 hours, with typical data shown in Fig. 2. Data obtained during the first ~10 hours of each accumulation measurement were disregarded to avoid inaccuracies from temperature stabilization time or equipment warm-up effects. Measurements began from either base chamber pressure, $<10^{-7}$ Torr, or by monitoring the pressure rate-of-rise starting from the pressure at the end of the previous run. Repeated measurements showed the measured outgassing rate was independent of the initial chamber pressure.

### *D. Coating analysis*

Test coupons coated with the TiN coating and the Silco coating were compared to a bare sample of the material used to fabricate the vacuum chambers. A Dektak-3-ST surface profilometer (2.5 µm radius diamond tip stylus, 3 mg tracking force) was used to measure surface roughness, and a JEOL JSM-6060 LV SEM equipped with EDS with a 20 kV electron beam provided 1,000 times magnification to evaluate surface morphology and chemical composition.

## III. RESULTS

### *A. Bare stainless steel*

The outgassing rates measured for the SS1 chamber are listed in Table I. The initial bake at 250 °C yielded an outgassing rate of 3.56 ($\pm0.0005$)$\times10^{-12}$ Torr L $s^{-1}$ $cm^{-2}$ (at 20°C), consistent with previous work at Jefferson Lab[2]. The extended heat treatment of the evacuated chamber at 400 °C in a hot air oven improved the outgassing rate by nearly a factor of 20, also consistent with past observations at Jefferson Lab and elsewhere[2,6-8,34].

To determine the effect of venting a chamber for assembly following a 400 °C heat treatment, the chamber was vented to air, re-evacuated, then baked first to 150 °C, and then at 250 °C with the outgassing rate evaluated after each bake. The 150 °C bake reduced the outgassing rate ~ 40% below that measured following the 400 °C bake. The reduction in outgassing rate due to the 150 °C bake is larger than would be expected due to hydrogen depletion through diffusion during the low temperature bake. Similar behavior seen elsewhere[13,35,36] is postulated to be a result of an oxide-layer diffusion barrier that forms after venting, evacuation and baking. The outgassing rate following the 250 °C bake increased to $2.79(\pm0.05)\times10^{-13}$ Torr L $s^{-1}$ $cm^{-2}$, higher than the rate following either the 400 °C or the 150 °C bake. This behavior has also been reported in literature[13,37] where it is believed that 250 °C provides sufficient thermal energy to liberate hydrogen atoms from strongly bound defect states. Once liberated, these hydrogen atoms serve to repopulate the reservoir of mobile hydrogen in the chamber walls, resulting in higher outgassing. A subsequent bake at 150 °C did not fully recover the best outgassing rate, but a final heat treatment at 360 °C for 150 h was effective in restoring the outgassing rate to its best measured value of $1.0(\pm0.5)\times10^{-13}$ Torr L $s^{-1}$ $cm^{-2}$, almost a factor of 40 lower than the initial outgassing rate of the system.

The outgassing rates following each bake were measured at room temperatures from 15 to 30 °C, shown on the Arrhenius plot in Fig. 3. The variation in outgassing rate with chamber temperature can be used to determine information about the thermally dependent hydrogen diffusion in the system [6,8,20]. The slope of the line in Fig. 3 yields the hydrogen diffusion activation energy, $E_D$, as shown in Eq. 2, where $\log A$ is a constant that is correlated to the initial concentration of hydrogen in the material, $R$ is the gas constant and $T$ is the chamber temperature in Kelvin.

$$\log q = \log A - \left(\frac{E_D}{R}\right)\frac{1}{T} \qquad (2)$$

For bare stainless steel, the activation energy calculation following the initial 250 °C bake yields a value of $E_D$ = 60.3 kJ/mol (= 14.4 kcal/mol = 0.63 eV) in agreement with results presented in literature[6,38], and will be discussed further in section IV.

## B. Amorphous Si coating

Amorphous silicon has been investigated as a surface coating to reduce pump-down times for systems in the high to ultra-high vacuum regime[26,27], but its applicability for baked UHV and XHV applications has not been evaluated. The a-Si chamber was coated prior to any heat treatment, while the SS1:a-Si chamber was coated following the extensive series of bakes and outgassing rate measurements described in the previous section for the SS1 chamber. The a-Si chamber was first baked at 90 °C, which is largely ineffective for removing water from bare stainless steel. The outgassing rate following this bake (see Table II) was a factor of two lower than the outgassing rate of the SS1 chamber following its initial 250 °C bake, which indicates that this a-Si coating provides a low outgassing rate following a bake at relatively low temperature.

However, subsequent bakes at 150 and 250 °C reduced the outgassing rate for the a-Si chamber only moderately (15 and 30% respectively). The heat treatment at 400 °C, which was effective in reducing outgassing in the SS1 chamber, instead increased the outgassing rate of the a-Si chamber by a factor of approximately two. This increase could arise from hydrogen diffusion from the steel into the coating during heat treatment, leading to a higher room temperature hydrogen concentration in the coating and resulting in higher outgassing rate, or alternately could be due to changes in the composition of the hydrogen compounds[39] ($SiH_x$, with x = 1, 2, 3) within the a-Si coating at temperatures above 350 °C. The chamber underwent additional 30 h bakes at 150 °C and then 250 °C. Following this lengthy bake history, the outgassing rate of the silicon-coated chamber was approximately the same as that obtained following the initial 90 °C bake. These results indicate that once a chamber is coated with amorphous silicon, heat treatments at temperatures up to 400 °C are ineffective at reducing the outgassing rate.

The next series of tests was designed to determine if the amorphous silicon coating would preserve a markedly low outgassing rate. The SS1 chamber that exhibited an outgassing rate of $1.0(\pm 0.5)\times 10^{-13}$ Torr L $s^{-1}$ $cm^{-2}$ was sent for coating with amorphous silicon. After coating, the chamber, now referred to as SS1:a-Si, was solvent cleaned and then baked at 90, 150, and 250 °C, with outgassing rates evaluated and compared to pre-coating values (see Table III). The 90 °C bake resulted in an outgassing rate of $1.004(\pm .005)\times 10^{-12}$ Torr L $s^{-1}$ $cm^{-2}$, similar to the 90 °C bake of the other a-Si coated chamber. Outgassing rates of 1.26 and $1.46(\pm 0.05)\times 10^{-13}$ Torr L $s^{-1}$ $cm^{-2}$ were obtained following bakes at 150 and 250 °C, and are increases of 26% and 46% respectively over the prior outgassing rate, which is significantly smaller than the ~280% increase in the bare steel outgassing rate following its 250 °C bake (4$^{th}$ in the series).

The results from both of the a-Si chambers suggest the coating is indeed hydrophobic compared to bare stainless steel, with outgassing rates in the low $10^{-12}$

Torr L $s^{-1}$ $cm^{-2}$ range obtained following bakes at only 90°C. The results also indicate that once the chamber has been coated with amorphous silicon, additional heat treatment does not reduce the outgassing rate, however a chamber that was heat treated prior to coating preserves the reduced outgassing rate following coating much better than uncoated steel.

The outgassing rates of the amorphous silicon-coated chambers were evaluated at a series of room temperatures (Fig. 4), and the decrease in outgassing rate with reduced temperature illustrates expected behavior associated with outgassing due to a thermally-dependent outgassing rate.

## *C. TiN coating*

The initial outgassing rate for the TiN coated chamber, measured following a 90 °C bake, yielded an excellent outgassing rate of $6.44(\pm.05)\times10^{-13}$ Torr L $s^{-1}$ $cm^{-2}$, which is lower than stainless steel following repeated 250 °C bakes, such as found in Ref. 2. The next three bakes in the series (150, 250 and 400 °C) had outgassing rates measured at various room temperatures, but these data did not follow the expected room temperature dependence. This is believed to be due to a leak in the isolation valve, so these data have been disregarded. After replacement of the isolation valve, a remarkably low outgassing rate of $4(\pm2)\times10^{-15}$ Torr L $s^{-1}$ $cm^{-2}$ was obtained following the 150 °C bake ($5^{th}$ in the series shown in Table IV), and this value decreased to $2(\pm20)\times10^{-16}$ Torr L $s^{-1}$ $cm^{-2}$ following the final bake at 250 °C. The room temperature dependence of these data sets followed the expected trend for thermally dependent outgassing behavior, with a decreased room temperature leading to a decrease in outgassing rate (see Fig. 5).

These remarkably low outgassing rates, though similar to some published values ($5.3\times10^{-15}$ and $1\times10^{-16}$ Torr L $s^{-1}$ $cm^{-2}$ from Refs. 20 and 23 respectively) require scrutiny.

The pressure within the TiN chamber remained near the bottom of the SRG measurement capability for days (pressure ~ $10^{-7}$ Torr). Repeated measurements of the outgassing rate, for times up to 200 hours reduced the statistical error in the measurements, but still leads to statistical and systematic error bars with magnitude similar to, or larger than, the measured value of the outgassing rate. If the pressure in the system were below the measurement capability of the SRG, the calculated outgassing rate could be dominated by a baseline drift of the gauge, though the noted room temperature dependence lends credibility to the measurements. It also must be noted that any pump speed in the system invalidates outgassing rate measurements because the accumulation method applies only to systems with no pumping. Elemental titanium is known to be effective in pumping gases such as hydrogen, and the presence of such pumping would yield an artificially low outgassing rate: it would require only 0.003 L/s pump speed and a typical outgassing rate of $1\times10^{-12}$ Torr L $s^{-1}$ $cm^{-2}$ to achieve apparent outgassing rates such as those found in this system.

## *D. Coating analysis*

Coating analysis was performed on one bare and two coated test coupons, all cut from the same material as the chambers, and with two coated by the manufacturers using the same parameters as during the test chamber coating processes. The bare stainless steel sample had an RMS roughness of 740 nm and the SEM image of this surface (Fig. 6a inset) shows characteristic roughness of a machined stainless surface as expected. Application of the a-Si coating served to smooth the surface, with 250 nm RMS roughness, and the SEM image of the coating appeared very uniform (Fig. 6b inset). The application of the TiN coating also served to smooth the bare stainless steel surface, with 510 nm RMS roughness, but the SEM image (Fig. 6c inset) shows obvious particulate matter on the surface. The particulate was also notable in the TiN coated chamber and was easily dislodged during handling, particularly after the bake. The surface roughness and coating thicknesses are summarized in Table V for bare, a-Si and TiN surfaces.

The EDS measurements made in conjunction with the SEM images illustrate the chemical composition of the three coated chambers (see Fig. 6). The EDS spectrum from the bare stainless steel coupon (Fig. 7a) shows the expected composition of 304L steel, with an additional carbon peak attributed to residual solvents from cleaning. The a-Si coated test coupon (Fig. 6b) shows strong peaks for both the underlying stainless steel and the silicon coating, consistent with the thin, inter-diffused silicon coating and the EDS technique, which is sensitive to the first ~ 1 micron of material. The EDS analysis of TiN (Fig. 6c) shows Ti and N peaks with an atomic ratio of approximately 1:1 (50.4% N and 49.6% Ti). However, careful EDS examination of smaller features in the TiN coating show inhomogeneity in the film, with Ti:N atomic ratios ranging from 0.58:1 to 1.17:1, suggesting the possible presence of elemental Ti in some areas. The TiN coating is much thicker than the a-Si coating (9 µm compared to 0.8 µm), thus the surface sensitive EDS measurements show minimal characteristics of the underlying stainless steel substrate.

# IV. DISCUSSION

The outgassing rate of a newly manufactured stainless steel vacuum system is initially governed by the rate at which hydrogen atoms diffuse through the chamber walls to the surface, where the concentration is depleted due to outgassing of hydrogen into the vacuum. This process, known as diffusion limited outgassing, is described by Fick’s diffusion equation. However, as the hydrogen concentration is reduced through repeated bakes or extended heat treatment, less hydrogen arrives at the surface per unit time due to diffusion. Eventually, the rate at which hydrogen atoms at the surface combine and desorb as $H_2$ molecules, known as recombination, becomes the dominating factor in the outgassing rate of the material rather than the diffusion rate to the surface[5,38-41].

Qualitatively, the transition from diffusion to recombination limited outgassing behavior can be seen on a graph of the outgassing rate as a function of heat treatment. The combination of time and temperature associated with a chamber's heat treatments can be described using the dimensionless Fourier number, $F_0$,

$$F_0 = 4D_H\left(\frac{t}{d^2}\right) \tag{3}$$

where $t$ is the processing time and $d$ is the thickness of the chamber wall. The temperature dependent diffusion constant, $D_H$, in Eq. 3 is related to the hydrogen diffusion activation energy in stainless steel, $E_D$, by the following expression:

$$D_H(T) = D_0 \exp\left(-E_D/RT\right) \tag{4}$$

where $D_0$ is termed the pre-exponential diffusion constant and assigned[42] a typical value of 0.012 $cm^2s^{-1}$ for 304L stainless steel, $R$ is the ideal gas constant and $T$ corresponds to temperature in Kelvin. The value of $E_D$ = 60.3 kJ/mol, determined in section IIIA for diffusion of hydrogen in bare steel after the initial 250 °C bake where bulk diffusion dominates the outgassing process, was used to determine $D_H(T)$ and $F_0$ for all systems.

The outgassing rate of each test chamber is plotted versus $F_0$ in Fig. 7. The straight line in Fig. 7 shows the modeled outgassing rate of a vacuum chamber based solely on diffusion per Fick's Law, yielding a straight line on a log scale, with outgassing rate falling continuously with increasing $F_0$.

The data for bare stainless steel support two regimes of outgassing. Initial bakes at 250 °C on the SS1 chamber and previously published data[2] (labeled as SS2 in Fig. 7) yield outgassing rates in good agreement with Fick's law for $F_0 < 1$ (see Fig. 7 inset). However, for $F_0 > 1$, the outgassing rates deviate from Fick's law, indicating that the outgassing rate has undergone the expected transition to recombination-limited behavior, with surface effects, rather than diffusion, dominating the outgassing rate process.

For the a-Si chamber, the graph of $q$ versus $F_0$ shows comparatively flat behavior. Recombination limited systems (such as the stainless steel chambers at $F_0 >> 1$) exhibit little change in $q$ versus $F_0$, suggesting that the outgassing rate for the a-Si chamber is strongly dominated by surface effects rather than by the rate of diffusion of hydrogen in the bulk stainless steel. The SS1:a-Si chamber was not heated to 400 °C following coating in order to preserve a known low outgassing rate for future experiments, and although the data are shown in the graph, dependence on $F_0$ is not clear from the limited data set.

The slope of the TiN outgassing rates vs. $F_0$ is quite steep, with the outgassing rate falling faster than Fick's law predicts due to diffusion from steel. Two possible explanations exist: either this TiN coating is an excellent diffusion barrier for hydrogen, and the bakes at even 90 and 150 °C are sufficient to deplete the hydrogen from the TiN coating itself, yielding exceptionally small outgassing rates, or alternatively the TiN coating may have a small pump speed, and the pump speed increases with additional heat treatments. A complementary experiment is underway using the TiN coated chamber, pumped by non-evaporable getter and ion pumps, to determine if the low apparent outgassing rate yields a markedly lower base pressure compared to that obtained using a bare stainless steel chamber.

## V. CONCLUSIONS

These studies used a bare stainless steel chamber to verify prior results and to validate the experimental setup for outgassing rate measurements of coated chambers. A 100 h bake at 400 °C reduced the outgassing rate of a bare stainless steel vacuum chamber from $3.560(\pm0.005)\times10^{-12}$ to $1.79(\pm0.05)\times10^{-13}$ Torr L $s^{-1}$ $cm^{-2}$. Subsequent bakes were far less effective at reducing the outgassing rate, indicating a transition from diffusion-limited, to recombination-limited outgassing behavior. Following venting to air and rebaking at 150 °C bake, the outgassing rate decreased modestly, but following a 250 °C bake, the outgassing rate increased by nearly a factor of 3 (from 1.01 to $2.79(\pm0.05)\times10^{-13}$ Torr L $s^{-1}$ $cm^{-2}$). These studies suggest that once low outgassing rates have been achieved, baking a stainless steel vacuum chamber at 250 °C in the recombination limited regime is detrimental to the outgassing rate.

This is the first investigation of the performance of amorphous-silicon coated chambers subjected to repeated bake cycles and heat treatments. The a-Si chamber had an outgassing rate of $2.355(\pm0.005)\times10^{-12}$ Torr L $s^{-1}$ $cm^{-2}$ following a 90 °C bake, which can typically only be reached with a 250 °C bake of a bare steel chamber, and is consistent with the hydrophobic nature of the surface advertised by the manufacturer. However, subsequent bakes at 150 and 250 °C reduced the outgassing rate minimally, and the 400 °C extended heat treatment increased the rate above the initial value obtained following the 90 °C bake, leading to the conclusion that heat treatment following coating is not effective in reducing the outgassing rate for a-Si coated chambers. Low outgassing rates were preserved for the chamber coated with a-Si after heat treatment (SS1:a-Si), indicating that coating a degassed chamber with a-Si can be beneficial for systems where a low outgassing rate is needed. Energy-dispersive x-ray spectroscopy data confirm the expected combination of silicon and 304L stainless steel in the spectrum.

The test results of the TiN coated chamber yielded the smallest outgassing rate of all the chambers: $6.44(\pm 0.05)\times 10^{-13}$ Torr L $s^{-1}$ $cm^{-2}$ following an initial 90 °C bake and $2(\pm 20)\times 10^{-16}$ Torr L $s^{-1}$ $cm^{-2}$ following the final bake in the series. The pressure, as measured by the SRG, remained at values near the gauge's measurement limit throughout the accumulation measurement. Thus, the uncertainty in the TiN outgassing rate is large and cannot preclude a small pumping effect of the TiN coating. Tests are underway to determine if this perceived low outgassing rate provides a comparable reduction in ultimate pressure in a vacuum chamber filled with pumps.

**Acknowledgements**

Notice: This work is authored by Jefferson Science Associates under U.S. DOE Contract No. DE-AC05-06OR23177 and with funding from the DOE R&D for Next Generation Nuclear Physics Accelerator Facilities Funding Opportunity Number: DE-FOA-0000339. The U.S. Government retains a non-exclusive, paid-up, irrevocable, world-wide license to publish or reproduce this manuscript for U.S. Government purposes. We thank Amy Wilkerson, Nick Moore and the College of William & Mary for the use of the profilometer to measure the surface roughness of the chamber materials.

# TABLES

**TABLE I:**

| Bake Temperature (°C) | Bake Time (hours) | Outgassing rate (Torr L $s^{-1}$ $cm^{-2}$) |
|---|---|---|
| 250 | 30 | $3.560(\pm 0.005)\times 10^{-12}$ |
| 400 | 100 | $1.79(\pm 0.05)\times 10^{-13}$ |
| 150 | 30 | $1.01(\pm 0.05)\times 10^{-13}$ |
| 250 | 30 | $2.79(\pm 0.05)\times 10^{-13}$ |
| 150 | 30 | $2.16(\pm 0.05)\times 10^{-13}$ |
| 360 | 150 | $1.0(\pm 0.5)\times 10^{-13}$ |

**TABLE II:**

| Bake Temperature (°C) | Bake Time (hours) | Outgassing rate (Torr L $s^{-1}$ $cm^{-2}$) |
|---|---|---|
| 90 | 30 | $2.355(\pm 0.005)\times 10^{-12}$ |
| 150 | 30 | $2.138(\pm 0.005)\times 10^{-12}$ |
| 250 | 30 | $1.556(\pm 0.005)\times 10^{-12}$ |
| 400 | 100 | $3.617(\pm 0.005)\times 10^{-12}$ |
| 150 | 30 | $2.806(\pm 0.005)\times 10^{-12}$ |
| 250 | 30 | $1.860(\pm 0.005)\times 10^{-12}$ |

**TABLE III:**

| Bake Temperature (°C) | Bake Time (hours) | Outgassing rate (Torr L $s^{-1}$ $cm^{-2}$) |
|---|---|---|
| 90 | 30 | $1.004(\pm 0.005)\times 10^{-12}$ |
| 150 | 30 | $1.26(\pm 0.05)\times 10^{-13}$ |
| 250 | 30 | $1.46(\pm 0.05)\times 10^{-13}$ |

**TABLE IV:**

| Bake Temperature (°C) | Bake Time (hours) | Outgassing rate (Torr L $s^{-1}$ $cm^{-2}$) |
|---|---|---|
| 90 | 30 | $6.44(\pm 0.05)\times 10^{-13}$ |
| 150 | 30 | not reliable |
| 250 | 30 | not reliable |
| 400 | 100 | not reliable |
| 150 | 30 | $4(\pm 2)\times 10^{-15}$ |
| 250 | 30 | $2(\pm 20)\times 10^{-16}$ |

**TABLE V:**

| Sample | Coating thickness | RMS roughness, nm |
|---|---|---|
| SS | None | 740 |
| a-Si | 800 nm | 250 |
| TiN | 10 µm | 510 |

## CAPTIONS

**TABLE I:** The uncoated chamber (SS1) bake history and corresponding outgassing rate at 20 °C.

**TABLE II:** The amorphous silicon coated (a-Si) chamber bake history and corresponding outgassing rates at 20 °C.

**TABLE III:** The chamber coated with amorphous silicon following heat treatment (SS1: a-Si) bake history and corresponding outgassing rate at 20 °C. The outgassing rate of SS1 prior to coating with a-Si was $1\times10^{-13}$ Torr L $s^{-1}$ $cm^{-2}$.

**TABLE IV:** The titanium nitride coated chamber (TiN) bake history and corresponding outgassing rate at 20 °C. Outgassing rates may be artificially low if there is any pump speed in the system, and the inhomogeneity found in the EDS analysis suggests the presence of some elemental Ti.

**TABLE V:** Coating thickness and RMS roughness values for bare steel, TiN coated and a-Si coated test samples.

FIGURE 1. Schematic of the outgassing measurement experimental apparatus during a 400 °C bake showing the chamber, SRG, isolation valve and pumps. For bakes at lower temperature, the first isolation valve, the ion pump and RGA were within the oven.

FIGURE 2. Representative rate-of-rise data at four room temperatures. These data were obtained using the SS1 stainless steel chamber, and correspond to outgassing rates between $1.4\times10^{-13}$ and $5.5\times10^{-13}$ Torr L $s^{-1}$ $cm^{-2}$ following a 150 °C bake (third in the series noted in Table I). A least-square linear fit was used for each data set to calculate outgassing rates.

FIGURE 3. The outgassing rates for the bare stainless steel (SS1) chamber as a function of inverse room temperature, with each data set obtained following a bake at the temperature noted in the legend. The error bars for statistical and systematic errors are smaller than the data points for these data. The slope yields the temperature dependent activation energy as described in the text.

FIGURE 4. The outgassing rates for the a-Si and the SS1:a-Si chambers as a function of inverse room temperature, with each data set obtained following a bake at the temperature noted in the legend. The low outgassing rate of the heat treated SS1 chamber (solid stars) was largely preserved following coating the chamber with a-Si and baking. The error bars for statistical and systematic error are comparable in size to the data points shown for these data.

FIGURE 5. The outgassing rates for the TiN-coated stainless steel chamber as a function of inverse room temperature, with each data set obtained following a bake at the temperature noted in the legend. The best outgassing rate for the SS1 bare stainless steel chamber is plotted for reference. Systematic and statistical errors, though small at larger outgassing rates, become significant at the lowest measured outgassing rates.

FIGURE 6. EDS and SEM data show coating composition and morphology for a) bare stainless steel, showing expected composition and morphology, b) a-Si on stainless steel, with a Si peak evident in conjunction with the steel substrate, and c) TiN coating with approximately 1:1 atomic ratio of Ti and N, but obvious particulate matter and pores visible across the surface. Inset SEM images each show an120×80 μm area.

FIGURE 7. Outgassing rates for each chamber are plotted vs. the Fourier number, $F_0$. The straight line represents the calculated outgassing rate for the systems using Fick's law, disregarding surface effects. For the steel chambers, the transition from diffusion limited reduction in outgassing rate to recombination limited behavior occurs near $F_0$ = 1, with the outgassing rates vs. $F_0$ in close agreement with Fick's law at low $F_0$, (see inset) and diverging from Fick's law toward recombination limited behavior at higher $F_0$ values. The a-Si chambers have outgassing rate largely independent of $F_0$, suggesting that the surface effects of the system dominate the outgassing, and diffusion of hydrogen from the material during bakes is not significant. The steep slope of the TiN indicates either an excellent diffusion barrier or a slight pumping speed in the coating that increases with additional heat treatment. Note: data labeled SS2 is from previously published data (Ref. 2).

FIGURE 1:

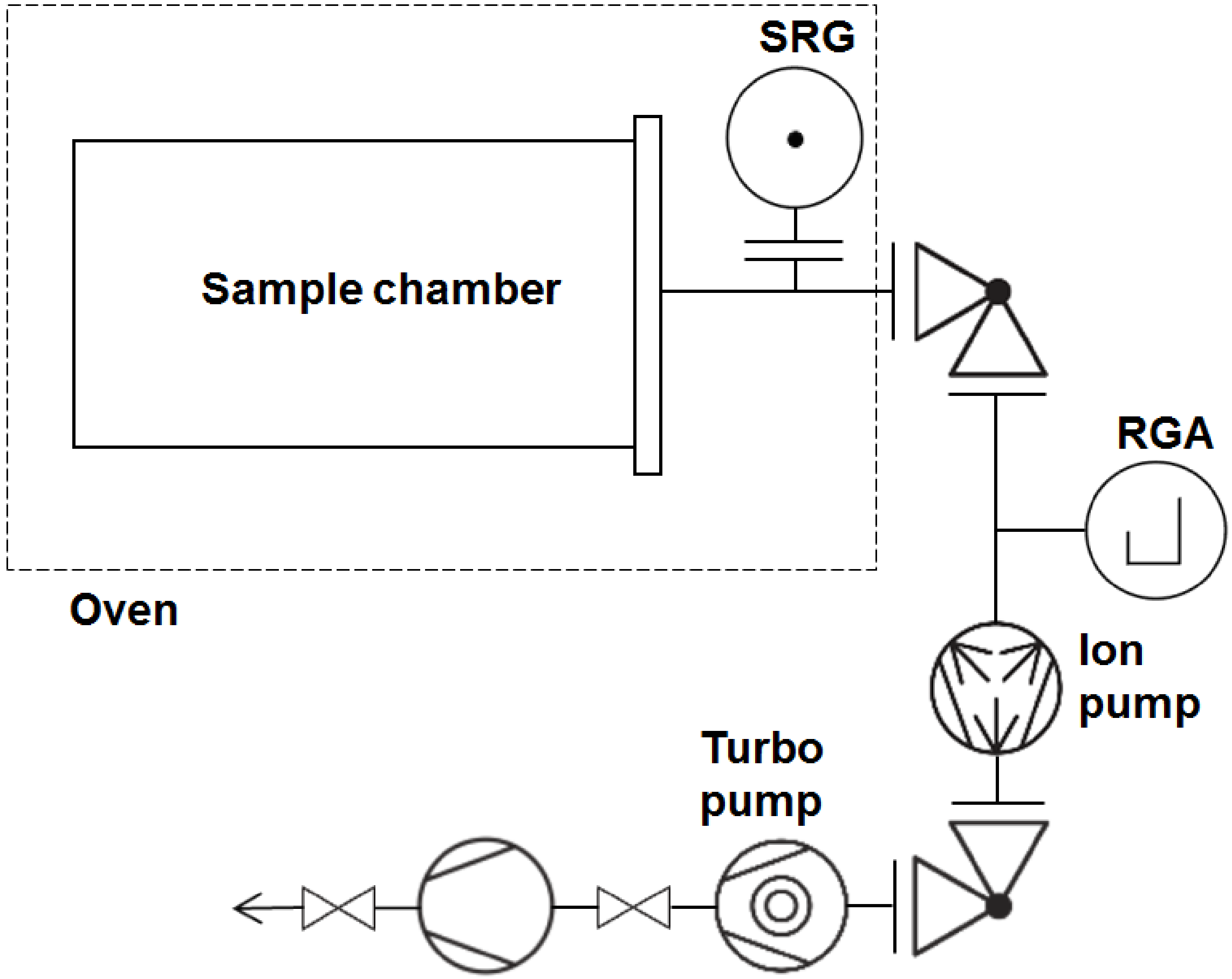

FIGURE 2:

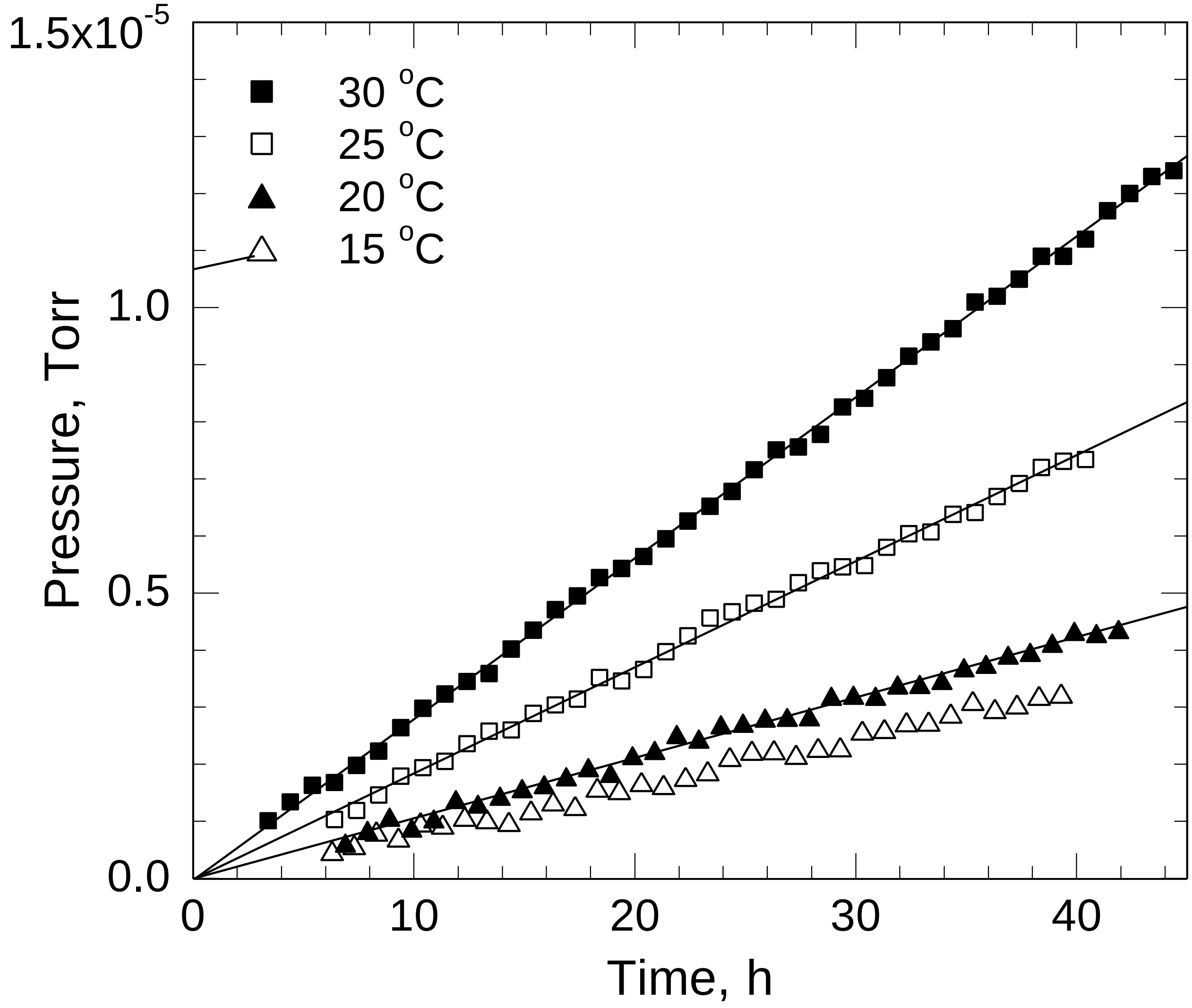

FIGURE 3:

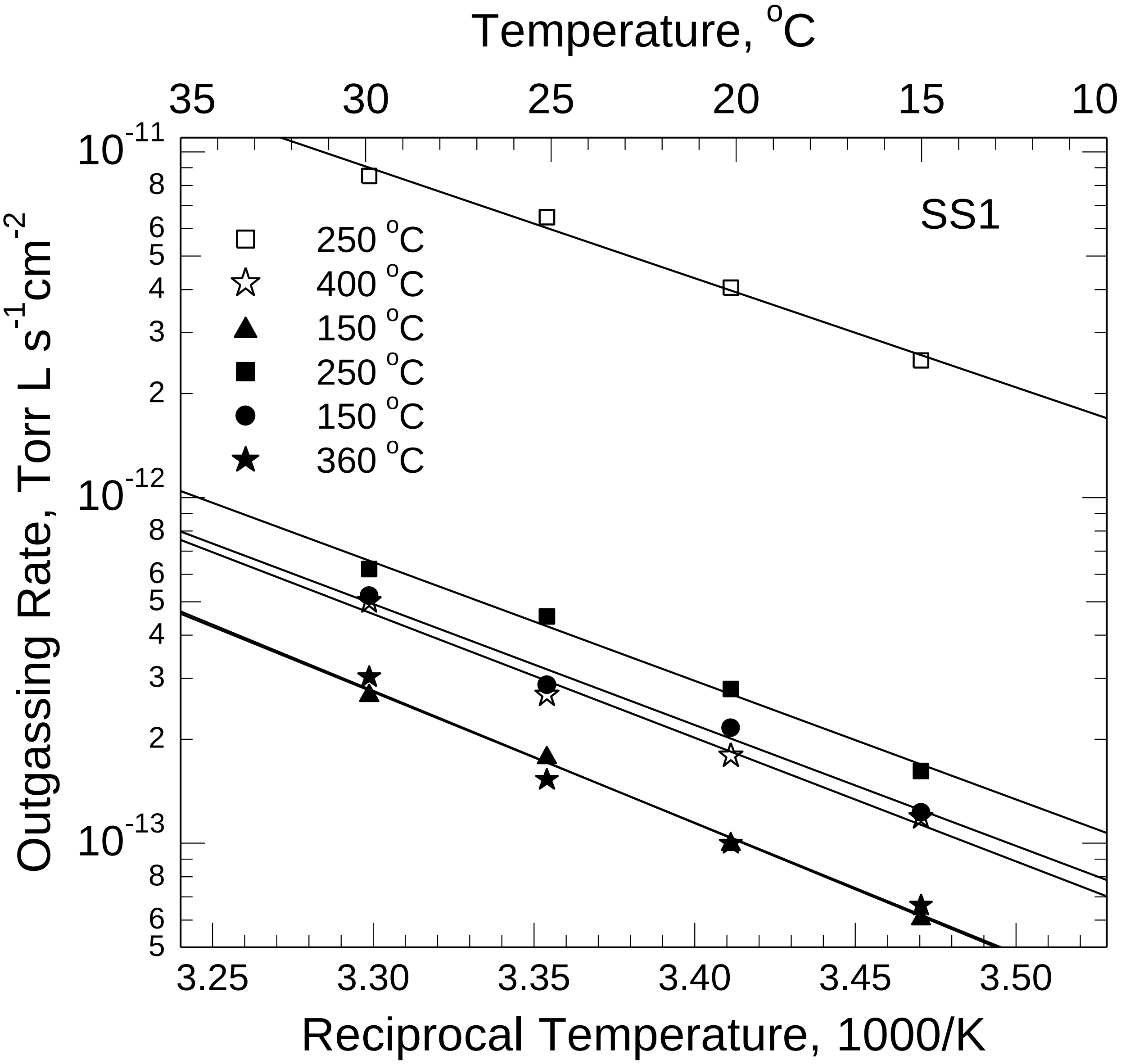

FIGURE 4:

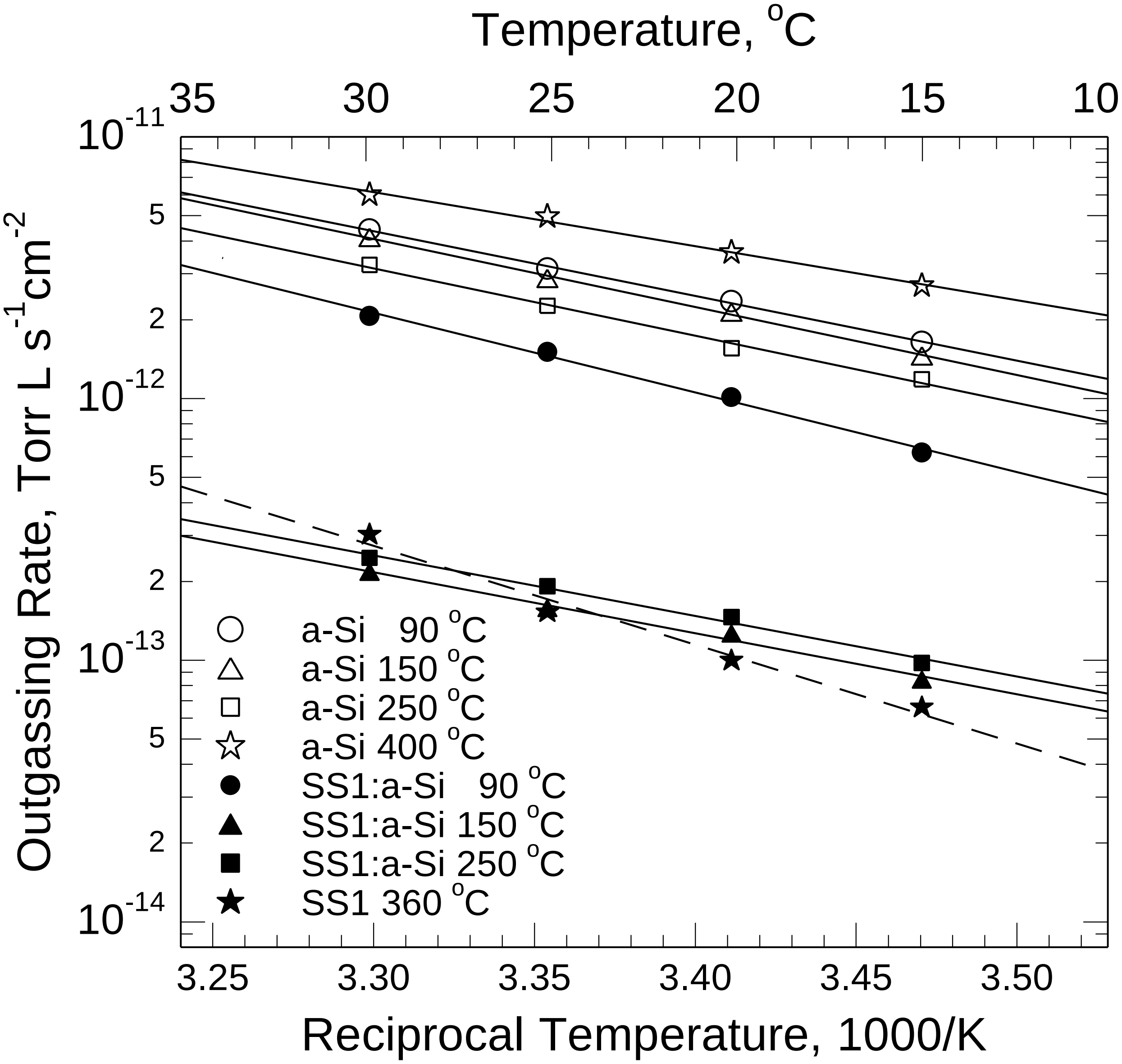

FIGURE 5:

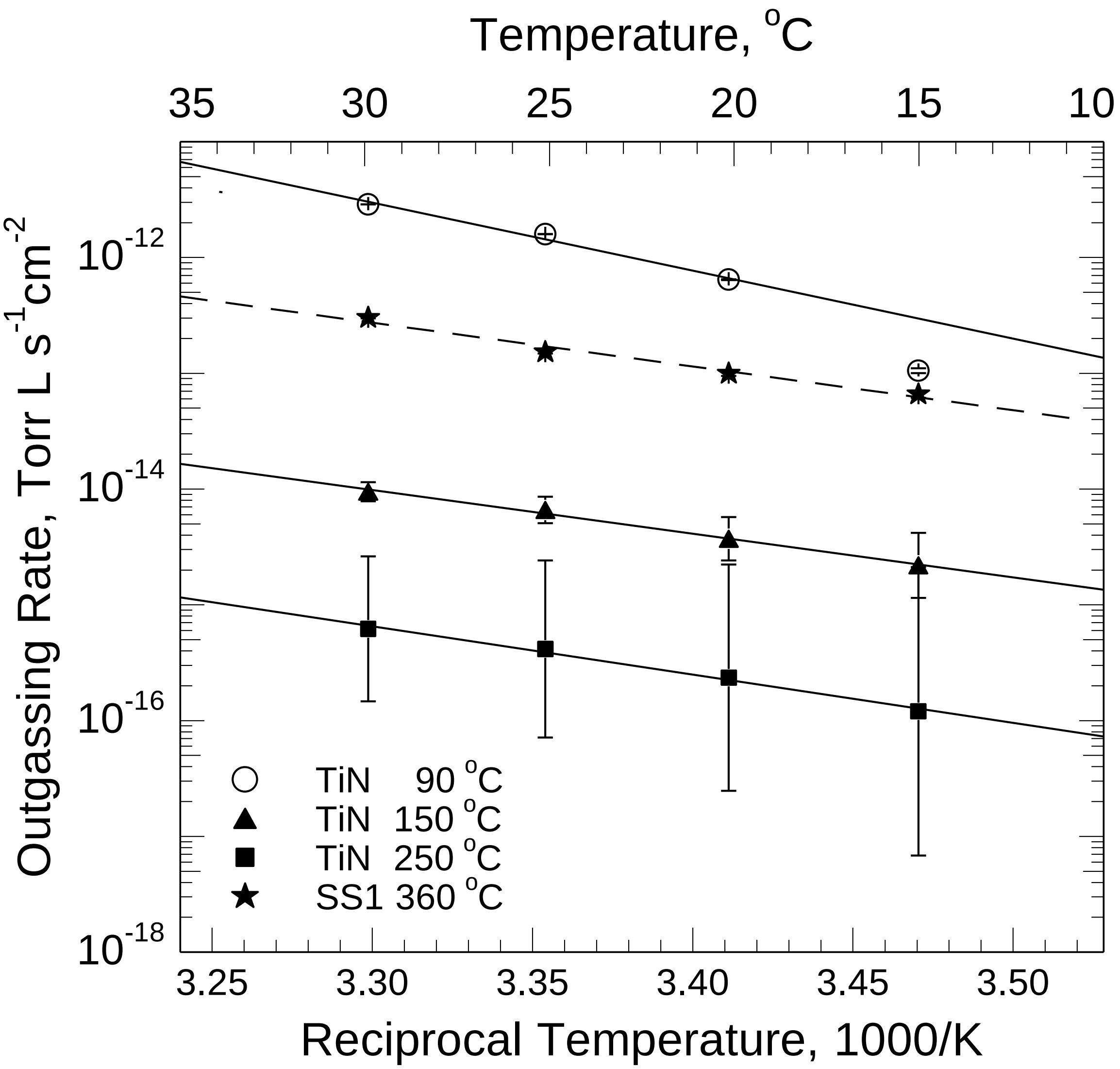

FIGURE 6:

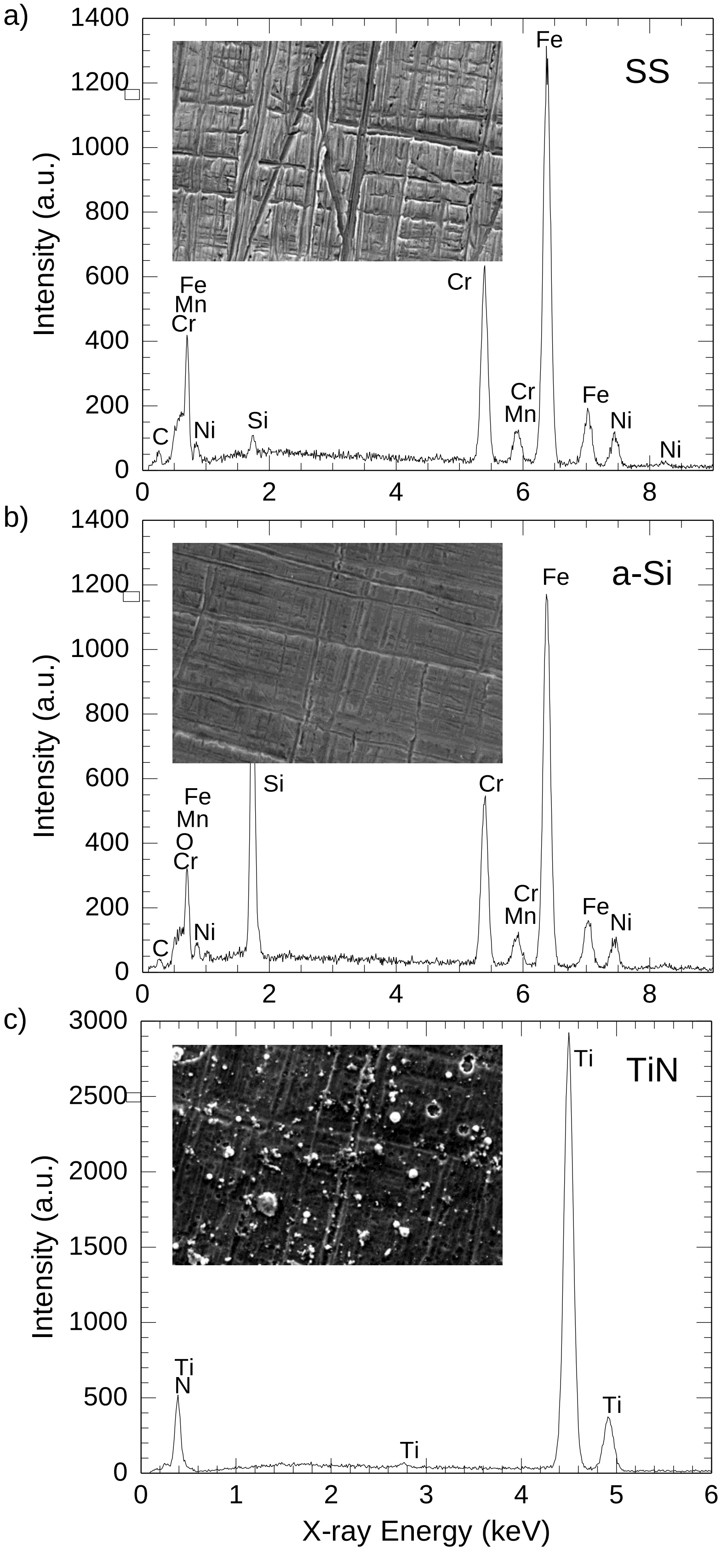

FIGURE 7:

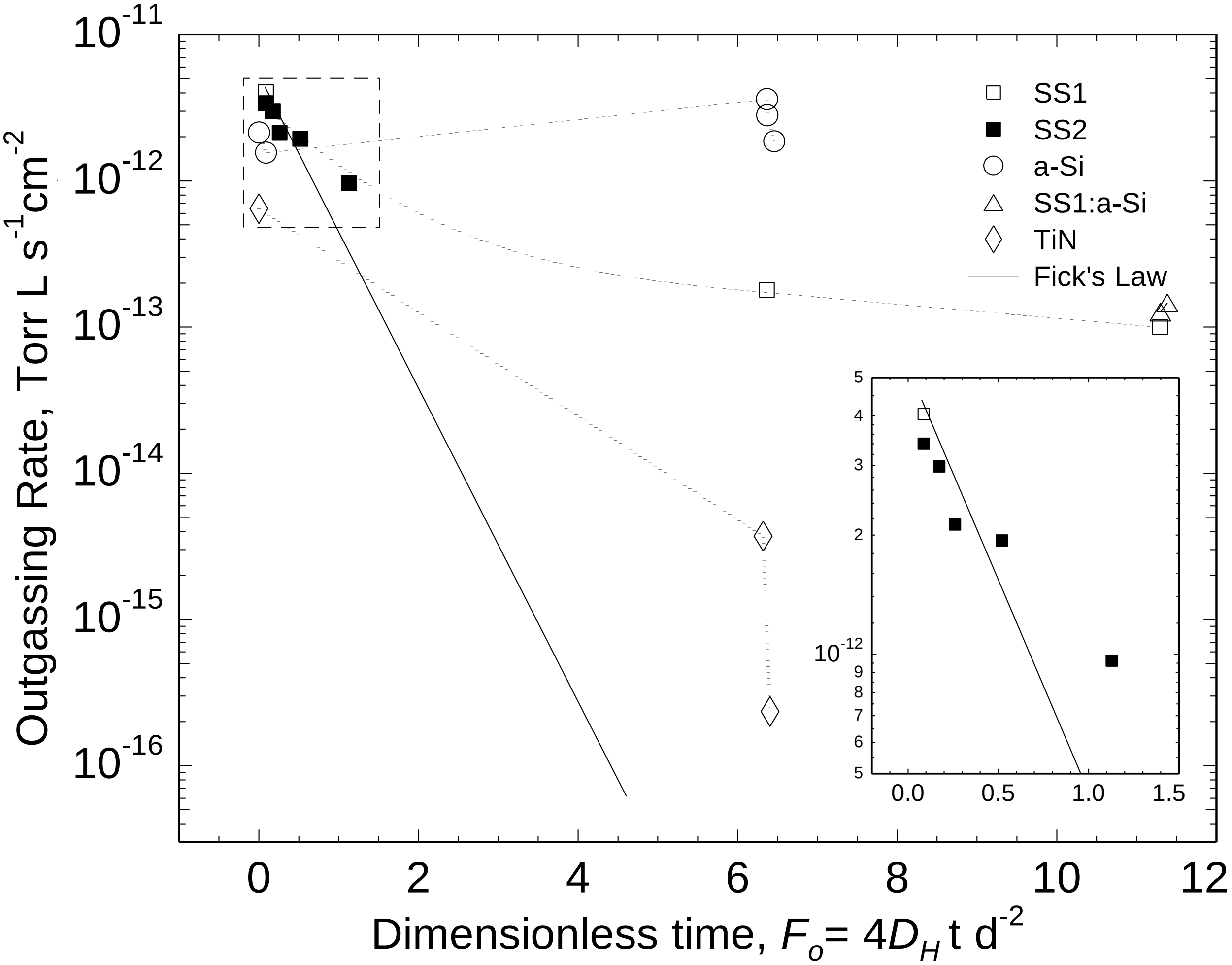